\documentclass[twocolumn,aps,pra,superscriptaddress,showpacs,tightenlines]{revtex4}
\usepackage{amsmath}
\usepackage{amsfonts}
\usepackage{graphicx}
\usepackage{epsfig}
\usepackage{color}
\usepackage[colorlinks,citecolor=blue]{hyperref}

\begin{document}

\title{Entanglement and quantum superposition induced by a single photon}
\author{Xin-You L\"{u}}
\email{xinyoulu@hust.edu.cn}
\affiliation{School of physics, Huazhong University of Science and Technology, Wuhan 430074, China}

\author{Gui-Lei Zhu}
\affiliation{School of physics, Huazhong University of Science and Technology, Wuhan 430074, China}

\author{Li-Li Zheng}
\affiliation{School of physics, Huazhong University of Science and Technology, Wuhan 430074, China}

\author{Ying Wu}
\email{yingwu2@126.com}
\affiliation{School of physics, Huazhong University of Science and Technology, Wuhan 430074, China}

\begin{abstract}
We predict the occurrence of single-photon-induced entanglement and quantum superposition in a hybrid quantum model, introducing an optomechanical coupling into the Rabi model. Originally, it comes from the photon-dependent quantum property of ground state featured by the proposed hybrid model. It associates with a single-photon-induced quantum phase transition, and is immune to the $A^2$ term of the spin-field interaction. Moreover, the obtained quantum superposition state is actual a squeezed cat state, which can give a significant precision enhancement in quantum metrology. This work offers an approach to manipulate entanglement and quantum superposition with a single photon, which might has potential applications in the engineering of new single-photon quantum devices, and also fundamentally broaden the regime of cavity QED. 
\end{abstract}
\maketitle
\section{Introduction}
Entanglement and quantum superposition, as the fundamental concepts of quantum mechanics, have wide applications in modern quantum technologies~\cite{Braunstein2005,Horodecki2009}. The ground-state entanglement and quantum superposition are usually in connection with the quantum phase transition (QPT) in the strongly correlated quantum systems~\cite{Osborne2002,Osterloh2002,Vidal2003,Lambert2004,Lambert2005,Ashhab2010}. Dicke model, describing a system of a quantized single-mode cavity field uniformly coupled to $N$ two-level systems, predict an equilibrium superradiant QPT in the thermodynamic limit, $N\rightarrow\infty$, i.e., the phase transition from a normal phase (NP) to a superradiant phase (SP) as increasing the spin-field coupling strength~\cite{Hepp1973}. Different from Dicke model, Rabi model considers a system of a quantized single-mode field (with frequency $\omega$) coupled to a single two-level system (with transition frequency $\Omega$), which is far from being in the thermodynamic limit. However, it is shown that the equilibrium superradiant QPT also exists in Rabi model, when the ratio of $\Omega$ to $\omega$ approaches infinity, i.e., the classical oscillator limit $\Omega/\omega\rightarrow\infty$ ~\cite{Hwang2015,Liu2017}. Associating with the superradiant QPT, the critical entanglement phenomenon~\cite{Lambert2004,Lambert2005} and quantum superposition of field~\cite{Ashhab2010} could be realized in cavity QED systems. However, they are limited by the so-called $A^2$ term of spin-field interaction, which corresponds to the debate on the existence of the equilibrium superradiant QPT in the cavity and circuit QED systems~\cite{Rzazewski1975,Knight1978,Nataf2010,Viehmann2011,Liberato2014,Vukics2014,Jaako2016}.

Recent advances in materials science and nano fabrication have led to spectacular achievements in the single photon technologies, including the single photon generation in cold atoms~\cite{Birnbaum2005,Peyronel2012}, quantum dots~\cite{Santori2002,Polyakov2011,Prechtel2013}, diamond color centers~\cite{Kennard2013}, or superconducting circuits~\cite{Pechal2014}, and the single photon detection based on quantum entanglement~\cite{Hempel2013} or cross-phase modulation~\cite{Grangier1998,Lukin2000,Xia2016}, etc.. These achievements have potential applications in quantum information science~\cite{Knill2001}, which leads to the recent explorations of single-photon transistor~\cite{Neumeier2013}, single-photon router~\cite{Zhou2013}, single-photon switch~\cite{Baur2014}, and single-photon triggered single-phonon source~\cite{Lodahl2016}. 

To combine the single photon technologies with entanglement and quantum superposition, here we investigate the ground state property of a hybrid quantum model, i.e., a Rabi model coupled to an ancillary cavity mode via a quadratic optomechanical coupling. Cavity optomechanics is a rapidly developing research field exploring the nonlinear photon-phonon interaction~\cite{Review1,Review2,Lu2015,Wu2015}. Typically the quadratic optomechanical coupling strength is very weak~\cite{Thompson2008,Bhattacharya2008}, which limits its application in quantum realm~\cite{liao2013}. Recent proposals have shown that it might be increased by a measurement-based method~\cite{Vanner2011}, the near-field effects~\cite{Li2012}, the good tunability of superconducting circuit~\cite{Kim2015}, or the modulation of photon-phonon interaction~\cite{Cirio2017}.

Interestingly, the concepts of single-photon-induced entanglement and quantum superposition are proposed in the hybrid quantum model. Physically, the proposed quantum model has a photon-dependent quantum property of ground state, which corresponds to a single-photon-induced superradiant QPT both in the cases of ignoring and including the $A^2$ term. This ultimately leads to the realizations of single-photon-induced entanglement and quantum superposition even in the weak coupling regime of spin-field interaction, and it is immune to the $A^2$ term. In general, the realizations of ground-state entanglement and quantum superposition in normal cavity QED systems are limited by the $A^2$ term. Moreover, here one can obtain a squeezed cat state of field, which could be used to enhance the detection precision in quantum metrology~\cite{Home2015,Knott2016}. As far as we know, this unconventional single-photon-induced entanglement and quantum superposition are identified for the first time, which is not only fundamental interesting, but can also inspire the engineering of new single-photon quantum devices.
\begin{figure}
\includegraphics[width=6.8cm]{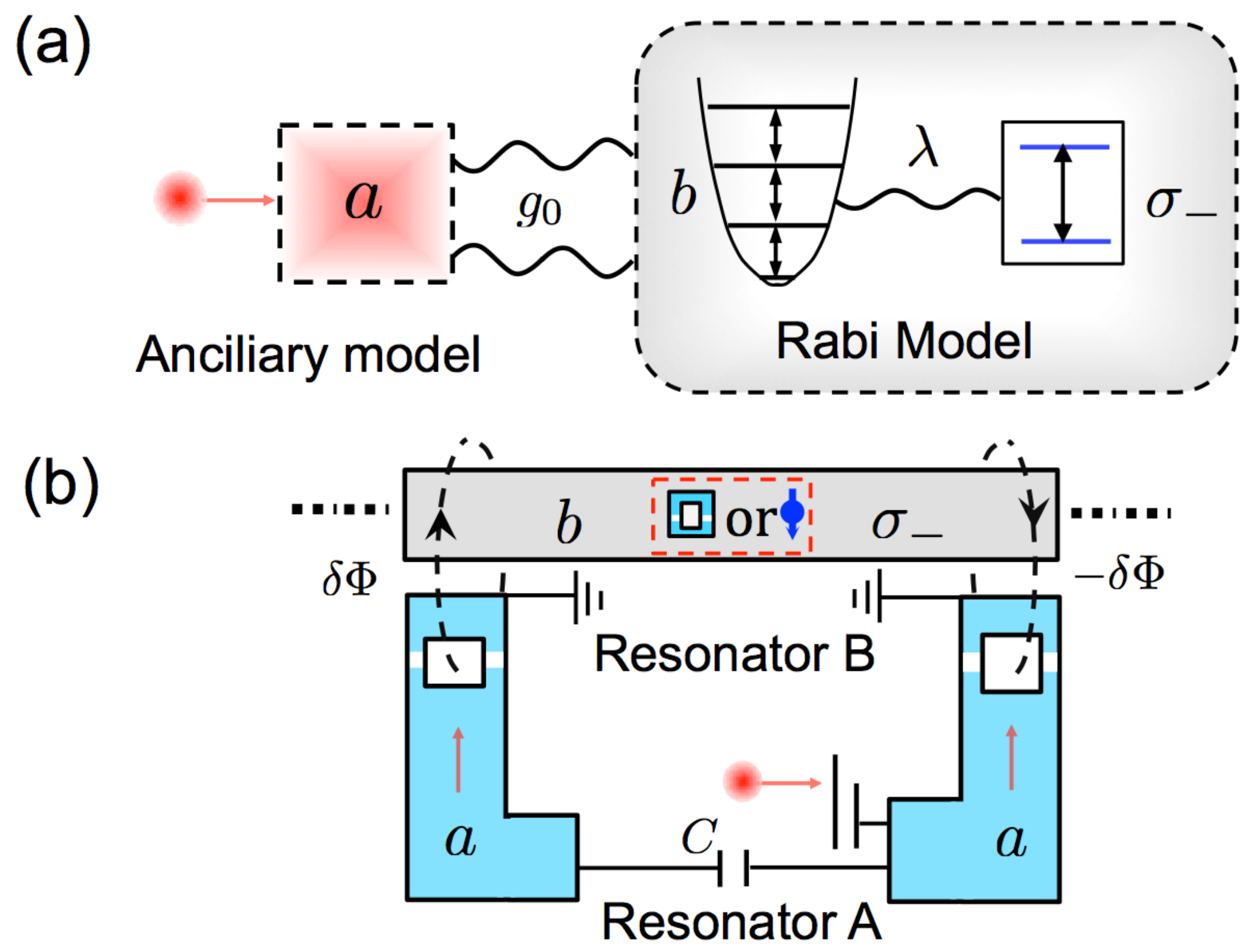}
\caption{(Color online) (a) A hybrid quantum model including a Rabi model quadratically coupled to an ancillary cavity mode $a$ with coupling strength $g_0$. (b) The implementation of this model in a superconducting circuit with the ability of simulating a quadratic optomechanical coupling~\cite{Kim2015} and coupling to the superconducting qubit or spin~\cite{Xiang2013}.}
\label{fig1}
\end{figure}

\begin{figure}
\centering
\includegraphics[width=0.43\textwidth]{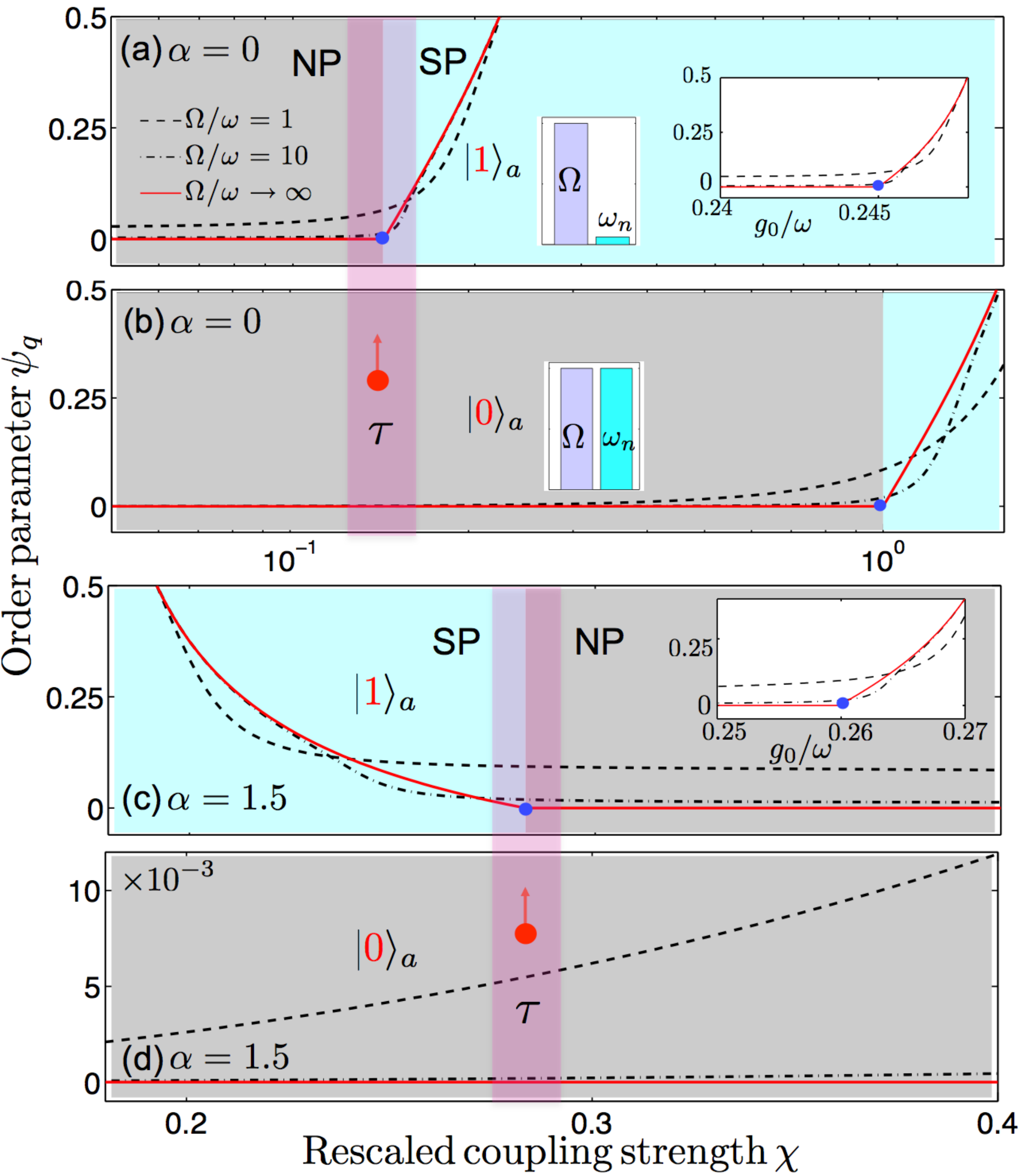}
\caption{(Color online) The order parameter $\psi_q$ versus $\chi$ (and $g_0/\omega$ in the inserted plots) for different $\Omega/\omega$ when (a,b) $\alpha=0$ and (c,d) $\alpha=1.5$. The inserted bar graphs present $\Omega/\omega_n$ when the mode $a$ is in $|0\rangle_a$, $|1\rangle_a$, and $\Omega=\omega$. The blue dots indicate the quantum critical point, where $\psi_q$ becomes finite from zero. The pink shading areas indicate the parameter range $\tau$ used to demonstrate single-photon-induced QPT. The system parameters are chosen as (a,b) $g_0/\omega=0.245$ and (c,d) $g_0/\omega=0.26$.}\label{fig2}
\end{figure}

\section{Model} 
We consider a hybrid quantum model depicted in Fig.$\,$\ref{fig1}(a) with total Hamiltonian ($\hbar=1$)
\begin{align}
H=H_{\rm an}+H_{\rm rm}- g_{0}a^{\dagger}a(b^{\dagger}+b)^2,\label{H_or}
\end{align}
where $a$ ($a^{\dagger}$) and $b$ ($b^{\dagger}$) are the annihilation (creation) operators of the ancillary cavity mode and the field mode of the Rabi model, respectively. Hamiltonian $H_{\rm rm}$ is given by $H_{\rm rm}=(\Omega/2)\sigma_{z}+\omega b^{\dagger}b-\lambda(b^{\dagger}+b)\sigma_{x}+(\alpha\lambda^2/\Omega)(b^{\dagger}+b)^2$, where $\sigma_{z}$ and $\sigma_{x}$ are the Pauli operators for the two-level system. It describes a two-level system $\sigma_-$ coupled to a field mode $b$ with coupling strength $\lambda$, and the $A^2$ term has been included in the last term. Normally $\alpha\geq1$ (decided by the Thomas-Reiche-Kuhn sum rule~\cite{Nataf2010}) corresponds to the case of implementing Rabi Model in cavity QED system, and $H_{\rm rm}$ is reduced to the Hamiltonian of a standard Rabi Model when $\alpha=0$. The ancillary cavity, with Hamiltonian $H_{\rm an}=\omega_a a^{\dagger}a$, quadratically couples to $b$ with coupling strength $g_0$~\cite{Bhattacharya2008}. This quadratic term provides a photon-dependent modification on the potential of field $b$. In the classical limit of the ancillary mode $a$, this model is approximately equivalent to the model studied in~\cite{Clerk2017}, where a time-dependent driving magnitude is employed. Here, we consider the case of $a$ being in a quantum state, i.e., Fock state $|n\rangle_a$, which allows the occurrence of single-photon-induced entanglement and quantum superposition. Moreover, the interplay between this quadratic interaction and the $A^2$ term is considered in our work, which makes our results immune to the $A^2$ term. Here we denote $|\!\!\uparrow\rangle$, $|\!\!\downarrow\rangle$ as the eigenstates of $\sigma_z$, and $|m\rangle_b$ as the eigenstate of $b^{\dagger}b$. Hamiltonian (\ref{H_or}) has $\mathbb{Z}_2$ symmetry associating with a well-defined parity operator $\Pi=e^{i\pi\mathcal{N}}$ (i.e., $[\Pi,H]=0$), where $\mathcal{N}=b^{\dagger}b+(1/2)(\sigma_z+1)$ is the total excitation number of system (excluding the ancillary cavity).

In principle, the proposed hybrid model could be realized in a quadratically coupled optomechanical system with a ``membrane-in-the-middle'' configuration. 
However the typical quadratic coupling in optomechanical system is too weak. It might be enhanced, by driving the mechanical system to large occupation numbers. But too strong driving will make the small displacement approximation used to derive the optomechanical interaction ineffective. Here we suggest to use the superconducting circuit depicted in Fig.\,\ref{fig1}(b) to implement our model. Specifically, as shown in Ref.\,\cite{Kim2015}, the coupling capacitor $C$ and the superconducting quantum interference devices (SQUIDs) forming resonator A offer an effective fixed semitransparent membrane and movable cavity ends, respectively. A relative displacement of the fixed membrane with respect to the center of resonator A is generated by synchronizing the motion of the moveable cavity ends, which is obtained by applying opposite flux variations $\pm\delta\Phi$ through the SQUIDs. Then the position quadrature of resonator B couples quadratically to the photon number of resonator A in a certain regime. Associating with the interaction between circuit cavity and the superconducting qubit or spin~\cite{Xiang2013}, our model could be realized in a superconducting circuit shown in Fig.\,\ref{fig1}(b), in which enough large $g_0/\omega$ might be reached in the further experiments by optimizing the coupling capacitance $C$, the bias flux through the SQUIDs and the geometrical arrangement of the circuit~\cite{Kim2015}.

\begin{figure}
\centering
\includegraphics[width=0.43\textwidth]{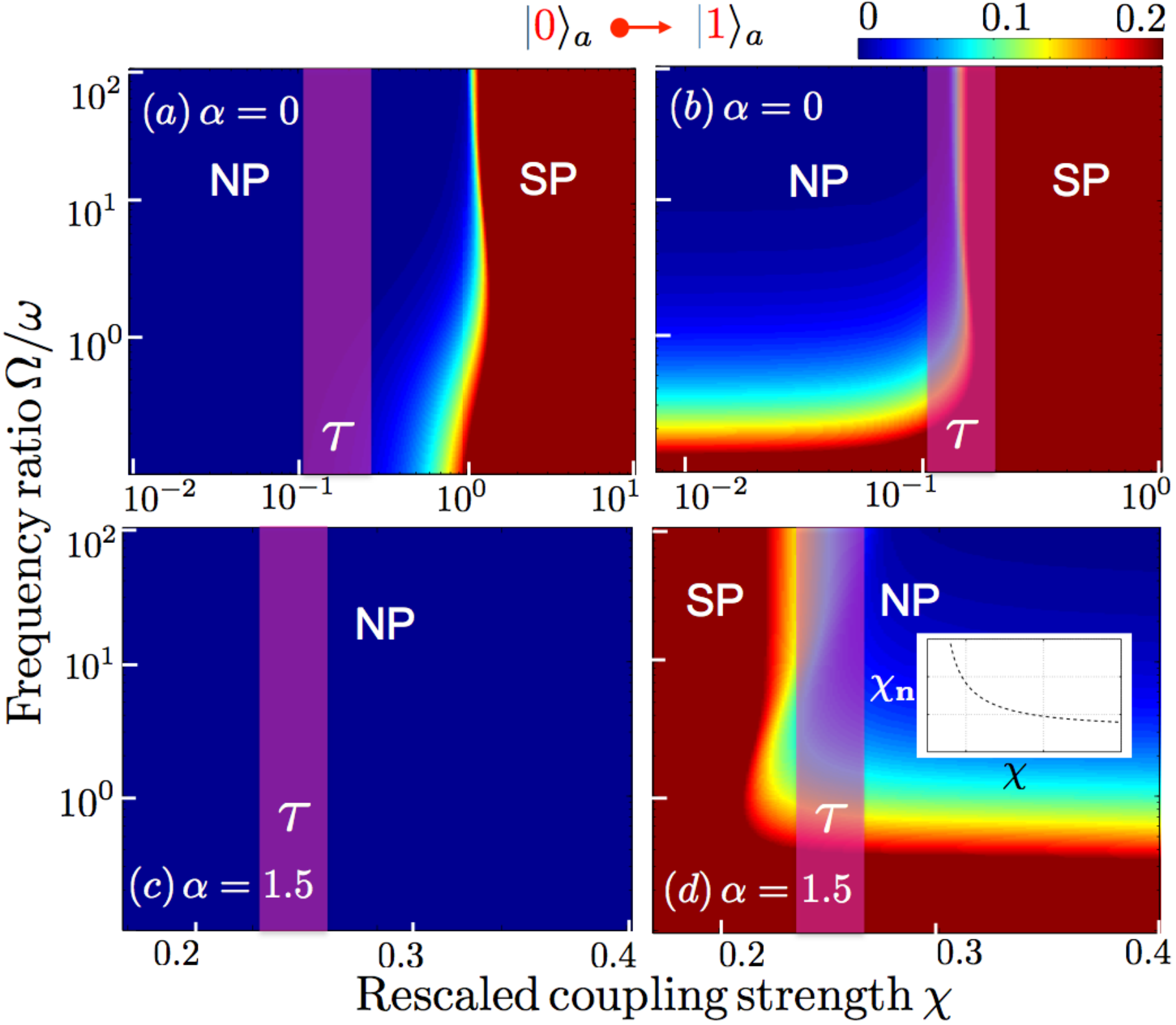}
\caption{(Color online) The order parameter $\psi_q$ versus $\chi$ and $\Omega/\omega$ when (a,b) $\alpha=0$, (c,d) $\alpha=1.5$. The subplots (a,c) and (b,d) correspond to mode $a$ being in Fock states $|0\rangle_a$ and $|1\rangle_a$, respectively. The insert of (d) present the dependence of $\chi_n$ on $\chi$, and the system parameters are same as Fig.\,\ref{fig2}.}\label{fig3}
\end{figure}

\section{Photon-dependent ground-state QPT} 
Considering the ancillary mode $a$ is prepared into a Fock state $|n\rangle_a$ ($n=0,1,...$), the number operator $a^{\dagger}a$ can be replaced by an algebraic number $n$. Then, applying a squeezing transformation $b=\cosh(r_n)b_n+\sinh(r_n)b_n^{\dagger}$ with $r_n=(-1/4)\ln[1+\alpha\chi^2-4ng_0/\omega]$ and a rescaled coupling strength $\chi=2\lambda/\sqrt{\Omega\omega}$, the Hamiltonian (\ref{H_or}) becomes
\begin{align}
H_{n}=\frac{\Omega}{2}\sigma_z+\omega_n b_n^{\dagger}b_n-\lambda_n(b_n^{\dagger}+b_n)\sigma_x+C_n,
\label{H_n}
\end{align}
where $\omega_n=\exp(-2r_n)\omega$, $\lambda_n=\exp(r_n)\lambda$ and $C_n=n\omega_{a}+[\exp(-2r_n)-1](\omega/2)$. It clearly shows that the proposed model is essentially equivalent to a photon-dependent Rabi model. 

In the $\Omega/\omega\rightarrow\infty$ limit (corresponding to $\Omega/\omega_n\rightarrow\infty$), Hamiltonian (\ref{H_n}) can be diagonalized analytically (see the Appendix A). A photon-dependent quantum critical point, $\chi_n=2\lambda_n/\sqrt{\Omega\omega_n}=1$ is obtained, corresponding to $\chi=\exp(-2r_n)=\sqrt{1+\alpha\chi^2-4ng_0/\omega}$ in term of the original system parameters. When $\chi<\exp(-2r_n)$ the system is in the NP, featured by an excitation energy $\omega_{e}$. The ground state of system is $|G\rangle_{\rm np}$, and it has a conserved $\mathbb{Z}_2$ symmetry (i.e., $\Pi|G\rangle_{\rm np}=|G\rangle_{\rm np}$), testified by the zero ground-state coherence of field $\langle b\rangle_g=0$. The excitation energy $\omega_{e}$ vanishes when $\chi=\exp(-2r_n)$, locating the superradiant QPT. When $\chi>\exp(-2r_n)$, the system enters into the SP and has an excitation energy $\tilde{\omega}_{e}$. Now the ground state of system becomes twofold degenerate, i.e., $|G\rangle^{\pm}_{\rm sp}$ (the detailed expression shown in Appendix A). It corresponds to a spontaneous $\mathbb{Z}_2$ symmetry breaking (i.e., $\Pi|G\rangle^{+}_{\rm sp}=|G\rangle^{-}_{\rm sp}$), as is evident from the non-zero ground-state coherence of field $\langle b\rangle^{\pm}_{g}=\pm \exp(r_n)\beta$. The rescaled ground-state occupation of field $b$, i.e., $\psi_{\rm q}=[\exp(-4r_n)\omega/\Omega]\langle b^{\dagger}b\rangle_g$, can be defined as the order parameter charactering this superradiant QPT. Because $\psi_{q}=0$ when $\chi<\exp(-2r_n)$, and $\psi_{\rm q}=(1/4)(\chi_n^2-\chi_n^{-2})$ becomes finite when $\chi>\exp(-2r_n)$, which is clearly displayed by the solid lines of Figs.\,\ref{fig2}(a-c). 

\begin{figure}
\centering
\includegraphics[width=0.43\textwidth]{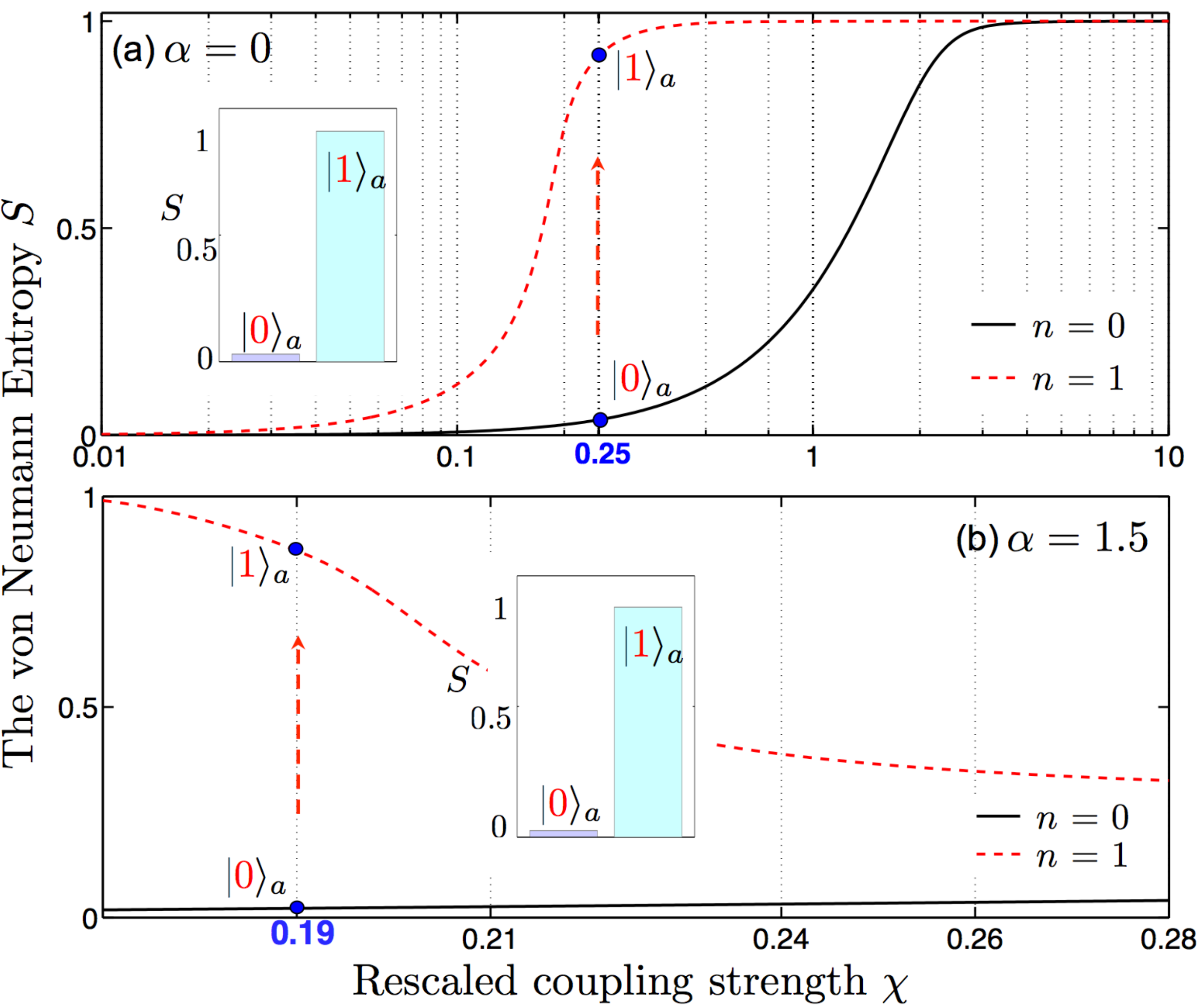}
\caption{(Color online) The von Neumann entropy $S$ versus $\chi$ for different $n$ when (a) $\alpha=0$ and (b) $\alpha=1.5$. The inserts indicate the values of $S$ corresponding to the blue dots. The red dashed arrows indicate the single-photon-induced quantum entanglement. The system parameters are same as Fig.\,\ref{fig2}.}
\label{fig4}
\end{figure}

\begin{figure*}
\centering
\includegraphics[width=0.9\textwidth]{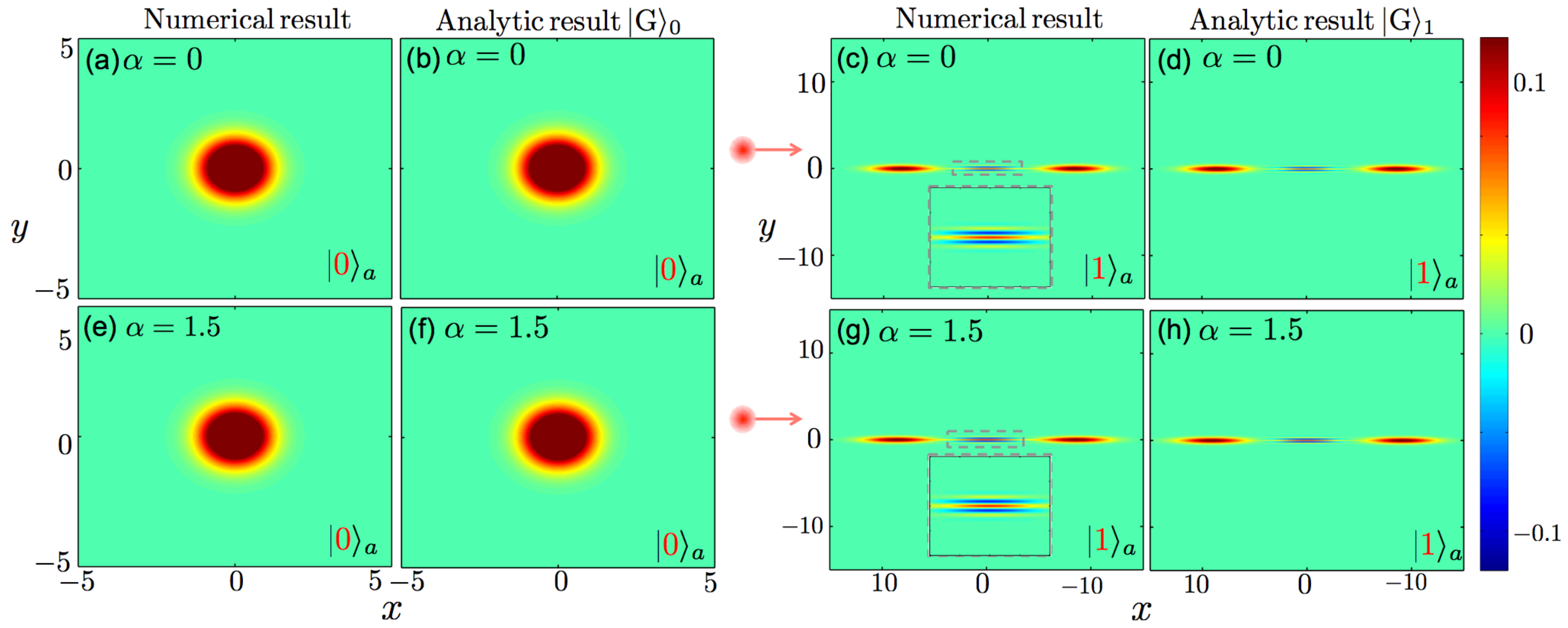}
\caption{(Color online) The Wigner function of the reduced density matrix $\rho_b$ when (a-d) $\alpha=0$ and (e-h) $\alpha=1.5$. The quadrature variables are $x=(b+b^{\dagger})/2$ and $y=-i(b-b^{\dagger})/2$. Corresponding to subplots (a,e,c,g), $\rho_b$ is obtained by diagonalizing the system Hamiltonian numerically and tracing out the qubit degree of freedom. The subplots (b,f) and (d,h) correspond to the approximate analytic ground state $|G\rangle_{0}$ and $|G\rangle_1$, respectively. The inserts of subplots (c,g) present the interfere fringe of cat state. The system parameters are same as Fig.\,\ref{fig2} except for the value of $\chi$ corresponding to the blue dots of Fig.\,\ref{fig4}.}\label{fig5}
\end{figure*}

Interestingly, the above photon-dependent quantum criticality leads to a single-photon-induced QPT, when we focus on the cases of $n=0,1$. Specifically, when the ancillary mode $a$ is in the vacuum state $|0\rangle_a$, Hamiltonian (\ref{H_n}) is reduced to a standard Rabi Hamiltonian. The superradiant QPT occurs at $\chi=1$ when $\alpha=0$, and it is prevented when $\alpha\geq1$ due to the no-go theorem~\cite{Hwang2015}. When $a$ is in the single-photon Fock state $|1\rangle_a$, the superradiant QPT occurs at $\chi=\exp(-2r_1)$, which could be much smaller than 1 for both the cases of $\alpha=0$ and $\alpha\geq1$, by properly choosing system parameters. Let's consider a parameter range $\chi\in\tau$ to check the occurrence of superradiant QPT ($\tau$ covering the single-photon-induced quantum critical point, $\chi=e^{-2r_1}$). As shown in Fig.\,\ref{fig2}, in the $\Omega/\omega\rightarrow\infty$, the superradiant QPT during $\tau$ is triggered by exciting a single photon in mode $a$ (i.e., $|0\rangle_a\rightarrow|1\rangle_a$). It corresponds to a single-photon-induced $\mathbb{Z}_2$ symmetry breaking, demonstrated by the ground-state coherence of field $\langle b\rangle_g$. Note that, this QPT describes the sudden change of ground state in a closed system as changing system parameter $\chi$ at zero temperature. It belongs to the equilibrium phase transition~\cite{Hepp1973,Hwang2015}, which is different from the non-equilibrium phase transition characterized by the steady state of the driven open systems~\cite{Dimer2007,Baumann2010,Baden2014}. 

Including the $A^2$ term, this superradiant QPT can still occur, since the parameter condition $\chi>\exp(-2r_1)$ can be satisfied even when $\alpha\geq1$. Moreover, the present superradiant QPT is reversed comparing with the case happened in a standard Rabi model~\cite{Hwang2015}, i.e., the transition from the NP to the SP occurs as decreasing the original system parameter $\chi$, as shown in Fig.\,\ref{fig2}(c). This originally comes from the competition between the quadratic term and the $A^2$ term in Hamiltonian (\ref{H_or}), which ultimately leads to the result that $\chi_n$ increases along with decreasing $\chi$ [see the insert of Fig.\,\ref{fig3}(d)]. 

In the finite $\Omega/\omega$ (corresponding to finite $\Omega/\omega_n$), the dependence of order parameter $\psi_q$ on $\chi$ (or $g_0/\omega$) clearly approaches the case of QPT occurring exactly with increasing $\Omega/\omega$ (see Figs.\,\ref{fig2} and \ref{fig3}). This tendency could be faster when $n=1$, comparing with the case of $n=0$. Physically, in our model, a single photon can induce a dramatically increasing of the value of $\Omega/\omega_n$ (see the bar graphs in Fig.\,\ref{fig2}). This leads to the result that, choosing the same value of $\Omega/\omega$, the case of $n=1$ can be closer to the $\Omega/\omega_n\rightarrow\infty$ limit, where the QPT occurs exactly. Here the results of finite $\Omega/\omega$ are obtained by numerically diagonalizing Hamiltonian (\ref{H_n}) in a large Hilbert space consisting of 1000 base vectors, and considering the squeezing transformation between modes $b$ and $b_n$. This Hilbert space also has been used in the following numerical calculations. 

\section{Single-photon-induced entanglement and quantum superposition} 
Figures\,\ref{fig2} and \ref{fig3} also show the approximate occurrence of superradiant QPT induced by a single photon in the finite $\Omega/\omega$, which leads to a single-photon-induced entanglement and quantum superposition.

Qualitatively, when the ancillary mode $a$ is in the vacuum state $|0\rangle_a$, Hamiltonian (\ref{H_n}) is reduced to a standard Rabi Hamiltonian. Under the conditions of $e^{-2r_1}<\chi\ll1$ and $\Omega\approx\omega$, the ground state of system is approximately as $|G\rangle_0=|0\rangle_b|\!\!\downarrow\rangle$, which is neither an entangled state nor a quantum superposition state. When the mode $a$ is in the single-photon Fock state $|1\rangle_a$, the frequency ratio $\Omega/\omega_n\gg1$ (see the bar graphs in Fig.\,\ref{fig2}), which allows the approximate occurrence of superradiant QPT when $\chi>e^{-2r_1}$. Correspondingly, the ground state of system approximately becomes $|G\rangle_1=(1/\sqrt{2})(|G\rangle^{+}_{\rm sp}+|G\rangle^{-}_{\rm sp})$  when $\chi>e^{-2r_1}$, which is a qubit-cavity entangled state. Moreover, from the ground state $|G\rangle_{1}$, we also could obtain the quantum superposition of the field $b$. One could measure the qubit in the $(|\!\!\downarrow\rangle_{+}\pm|\!\!\downarrow\rangle_{-})/\sqrt{2}$ basis (the definition of  $|\!\!\downarrow\rangle_{\pm}$ is shown in Appendix A). Depending on the outcome of the measurement, the state of the field $b$ is approximately projected into one of the following squeezed cat states
\begin{align}
|\Psi\rangle_1^{\rm sup} = \frac{1}{\sqrt{2}}S(\tilde{r}_{\rm tot})\left[D(|\beta|)|0\rangle_b \pm D(-|\beta|)|0\rangle_b\right].
\end{align}

Quantitatively, to show the above single-photon-induced quantum entanglement more clearly, we numerically calculate the von Neumann entropy $S=-{\rm tr}(\rho_b{\rm log}_2\rho_b)$ of the reduced density matrix $\rho_b$ of the field mode, and present the dependence of $S$ on $\chi$ in Fig.\,\ref{fig4}. It is shown that strong qubit-field quantum entanglement is triggered by injecting a single photon into the ancillary cavity, i.e., $|0\rangle_a\rightarrow|1\rangle_a$. This single-photon-induced quantum entanglement is immune to the $A^2$ term [see Fig.\,\ref{fig4}(b)]. Moreover, this strong qubit-field entanglement could be realized in a relatively weak coupling regime, i.e., $\chi\ll1$. However, in the normal Rabi Model (see the case of $n=0$ corresponding to the black solid lines in Fig.\,\ref{fig4}), the realization strong qubit-field entanglement requires the ultrastrong-coupling regime $\chi>1$ and ignoring the $A^2$ term.

To show the single-photon-induced quantum superposition, in Fig.\,\ref{fig5}, we present the Wigner function of the reduced density matrixes $\rho_b$ with the numerical results and the approximate analytic ground states $|G\rangle_0$, $|G\rangle_1$, respectively. First of all, it clearly presents that the single-photon-induced quantum superposition can be realized both in the cases of ignoring and including the $A^2$ term. Here the quantum superposition state is actually a squeezed cat state, as shown in Figs.\,\ref{fig5}(c,d,g,h). Secondly, comparing the exactly numerical results [i.e., Figs.\,\ref{fig5}(a,e) and (c,g)] with the analytic results [i.e., Figs.\,\ref{fig5}(b,f) and (d,h)], it is shown that the analytic ground states $|G\rangle_{0}$ and $|G\rangle_1$ can represent the system ground states with high fidelity. Then one could obtain a squeezed cat state $|\Psi\rangle^{\rm sup}_1$ with high fidelity after doing the qubit measurement into the ground state of system $|G\rangle_{1}$.

\section{Conclusions} 
In conclusion, we have proposed a hybrid quantum model, which is equivalent to a photon-dependent Rabi model. Interestingly, this hybrid quantum model allows the occurrence of single-photon-induced entanglement and quantum superposition. We also showed that these single-photon-induced quantum property will not be limited by the so-called $A^2$ term. Moreover, here the obtained quantum superposition state induced by a single-photon actually is a squeezed cat state, which has potential applications in quantum metrology~\cite{Home2015}. This work may offer the prospect of exploring the single-photon-induced ground-state quantum property together with its applications in the high-precision single-photon quantum technologies.

{\it Note added.---} During the final stages of this paper, two related works by Clerk's group~\cite{Clerk2017} and Nori's group~\cite{Qin2018} appeared. appeared.

\begin{acknowledgements}
This work is supported by the National Key R \& D Program of China (Grant No. 2016YFA0301203) and the National Science Foundation of China (Grant Nos. 11374116, 11574104 and 11375067).
\end{acknowledgements}

\appendix
\section{Diagonalization procedure of Hamiltonian (\ref{H_n})}
According to the diagonalization procedure used in Ref.\,\cite{Hwang2015}, we can diagonalize Hamiltonian (\ref{H_n}) in the $\Omega/\omega_n\rightarrow\infty$ limit (corresponding to $\Omega/\omega\rightarrow\infty$ limit in terms of the original system parameters). 

Specifically, when $\chi<\exp(-2r_n)$ (corresponding to the NP), Hamiltonian (\ref{H_n}) can be diagonalized according to the following procedure. Applying a unitary transformation $U^{\dagger}H_{n}U$ with 
\begin{align}
U=\exp(S)=\exp\left[\frac{\lambda_n}{\Omega}(b_n+b_n^{\dagger})(\sigma_{+}-\sigma_{-})\right],
\label{es1}
\end{align}
we obtain 
\begin{align}
H'_n=U^{\dagger}H_{n}U=\sum^{\infty}_{j=0}\frac{1}{j!}[H_{n},S]^{(j)},
\label{es2}
\end{align}
where the commutation rule is defined as $[H_n,S]^{(j)}\equiv[[H_n,S]^{(j-1)},S]$ with $[H_n,S]^{(0)}=H_n$. Expanding Eq.\,(\ref{es2}), we can obtain
\begin{align}
H'=&\frac{\Omega}{2}\sigma_z+\omega_nb^{\dagger}b+\frac{\chi^2_n\omega_n}{4}(b^{\dagger}_{n}+b_n)^2\sigma_z+C_n\nonumber
\\
&+\frac{\chi_n\omega_n}{2}\sqrt{\frac{\omega_n}{\Omega}}(b_n^{\dagger}-b_n)(\sigma_{+}-\sigma_{-})\nonumber
\\
&+\frac{\chi_n^3\omega_n}{6}\sqrt{\frac{\omega_n}{\Omega}}(b_{n}+b^{\dagger}_n)^3\sigma_x+\mathcal{O}(\sqrt{\frac{\omega_n}{\Omega}}),
\label{es3}
\end{align}
where the last term denotes the high-order terms of $\sqrt{\omega_n/\Omega}$. In the limit $\Omega/\omega_n\rightarrow\infty$ (originally from $\Omega/\omega\rightarrow\infty$), the fifth, sixth terms of Eq.\,(\ref{es3}) and the high order terms of $\sqrt{\omega_n/\Omega}$ [i.e., the last term of Eq.\,(\ref{es3})] become zero. Then, projecting the Hamiltonian into the spin down subspace, the system Hamiltonian becomes   
\begin{align}
\label{es4}
H_{\rm np}=\,&\omega_nb_n^{\dagger}b_n-\frac{\chi^2_n\omega_n}{4}(b^{\dagger}_{n}+b_n)^2-\frac{\Omega}{2}+C_n.
\end{align}
This Hamiltonian could be diagonalized to $H_{\rm np}=\omega_e e^{\dagger}e+E_g$ by a squeezing transformation $b_n=e\cosh(l)+e^{\dagger}\sinh(l)$ with a squeezing parameter $l=-(1/4)\ln(1-\chi^2_n)$. Here the excitation energy $\omega_e$ and the ground state energy $E_g$ are given by
\begin{subequations}
\begin{align}
\label{es5}
\omega_e&=\omega_n\sqrt{1-\chi^2_n},
\\
E_g&=\frac{\omega_n}{2}(\sqrt{1-\chi^2_n}-1)-\frac{\Omega}{2}+C_n.
\end{align}
\end{subequations}
The corresponding ground state of system is $|G\rangle_{\rm np}=S(r_{\rm tot})|0\rangle_b|\!\downarrow\rangle$ with $S(r_{\rm tot})=\exp[r_{\rm tot}(b^{\dagger2}-b^2)/2]$ and $r_{\rm tot}=r_n+l$. This ground state has a conserved $\mathbb{Z}_2$ symmetry (i.e., $\Pi|G\rangle_{\rm np}=|G\rangle_{\rm np}$), testified by the zero ground-state coherence of field $\langle b\rangle_g=0$. 

The excitation energy $\omega_e$ is real only for $\chi\leq\exp(-2r_n)$ (corresponding to $\chi_n\leq1$) and vanishes when $\chi=\exp(-2r_n)$, locating the occurrence of superradiant QPT. When $\chi>\exp(-2r_n)$, the system enters into the SP, and Eq.\,(\ref{es4}) becomes invalid due to the field $b_n$ is macroscopically occupied (being proportional to $\Omega/\omega_n$). In this case, we firstly displace the field mode $b_n$ with an amplitude $\beta=\pm\sqrt{\frac{\Omega}{4\omega_n}(\chi^2_n-\chi^{-2}_n)}$ (i.e., $b_{n}\rightarrow\tilde{b}_n+\beta$), and then the system Hamiltonian becomes
\begin{align}
\label{es6}
\!\!\!\!\tilde{H}_{n}=\omega_n\tilde{b}_n^{\dagger}\tilde{b}_n+\frac{\tilde{\Omega}}{2}\tilde{\sigma}_{z}-\tilde{\lambda}(\tilde{b}_n^{\dagger}+\tilde{b}_n)\tilde{\sigma}_{x}+\omega_n\beta^2+C_n,
\end{align}  
where the rescaled system coefficients $\tilde{\Omega}=\chi^2_n\Omega$, $\tilde{\lambda}=\sqrt{\Omega\omega_n}/(2\chi_n)$. Here $\tilde{\sigma}_z$, $\tilde{\sigma}_x$ are the redefined Pauli operators in the rotated spin eigenstates given by
\begin{subequations}
\label{es7}
\begin{align}
\tilde{|\!\downarrow\rangle}=\cos\theta|\!\downarrow\rangle-\sin\theta|\!\uparrow\rangle,
\\
\tilde{|\!\uparrow\rangle}=\sin\theta|\!\downarrow\rangle+\cos\theta|\!\uparrow\rangle,
\end{align} 
\end{subequations}
and $\tan(2\theta)=-4\lambda_n\beta/\Omega$. Note that Hamiltonian (\ref{es6}) has the same formation as Hamiltonian (\ref{H_n}). Then, by employing the similar procedure used to derive $H_{\rm np}$, Hamiltonian (\ref{es6}) can be diagonalized to $H_{\rm sp}=\tilde{\omega}_e\tilde{e}^{\dagger}\tilde{e}+\tilde{E}_g$ with 
\begin{subequations}
\label{es7}
\begin{align}
\tilde{\omega}_e&=\omega_n\sqrt{1-\chi_n^{-4}},
\\
\tilde{E}_g&=\frac{\omega_n}{2}(\sqrt{1-\chi^{-4}_n}-1)-\frac{\Omega}{4}(\chi^2_n+\chi^{-2}_n)+C_n.
\end{align}
\end{subequations}
Here the introduced operator $\tilde{e}$ is decided by a squeezing transformation $\tilde{e}=\tilde{b}_n\cosh(\tilde{l})-\tilde{b}^{\dagger}_n\sinh(\tilde{l})$ with a squeezing parameter $\tilde{l}=-(1/4)\ln(1-\chi^{-4}_n)$. Now the ground state of system becomes twofold degenerate given by $|G\rangle^{\pm}_{\rm sp}=D_n(\pm|\beta|)S(\tilde{r}_{\rm tot})|0\rangle_b|\!\downarrow\rangle_{\pm}$ with $D_n(\beta)=\exp(\beta b^{\dagger}_n-\beta^*_rb_n)$, $\tilde{r}_{\rm tot}=r_n+\tilde{l}$, and the spin states $|\!\downarrow\rangle_{\pm}$ given by
\begin{subequations}
\begin{align}
|\!\downarrow\rangle_{\pm}=\frac{\sqrt{1+\chi^{-2}_n}}{2}|\!\downarrow\rangle\pm\frac{\sqrt{1-\chi^{-2}_n}}{2}|\!\uparrow\rangle.
\end{align} 
\end{subequations}
Consequently, the $\mathbb{Z}_2$ symmetry of this ground state is spontaneously broken (i.e., $\Pi|G\rangle^{+}_{\rm sp}=|G\rangle^{-}_{\rm sp}$), testified by the non-zero ground-state coherence of field $\langle b\rangle^{\pm}_{g}=\pm e^{r_n}|\beta|$. 

To character this superradiant QPT more clearly, the rescaled ground-state occupation of field $b$ could be defined as the order parameter, i.e., $\psi_{\rm q}=[e^{-4r_n}\omega/\Omega]\langle b^{\dagger}b\rangle_g$. Based on the obtained ground state in the normal phase $|G\rangle_{\rm np}$ and the superradiant phase $|G\rangle^{\pm}_{\rm sp}$, we analytically calculate this order parameter, and obtain that $\psi_q=0$ for $\chi<\exp(-2r_n)$ (corresponding to the NP) and $\psi_q=(1/4)(\chi^2_n-\chi^{-2}_n)$ becomes finite for $\chi>\exp(-2r_n)$ (corresponding to the SP). This property ensures the validity of the defined $\psi_{\rm q}$ as an order parameter for charactering the QPT.

\end{document}